\documentclass[prl,twocolumn,showpacs,amsmath,amssymb,superscriptaddress]{revtex4}
\usepackage{pifont,amssymb,epsf,graphicx}
\begin{document}
\title{Nonmagnetic Impurity Resonances as a Signature of Sign-Reversal Pairing in the FeAs-based Superconductors }
\author{Degang Zhang}
\affiliation{Texas Center for Superconductivity and Department
of Physics, University of Houston, Houston, TX 77204, USA}

\begin{abstract}
The low energy band structure of the FeAs based superconductors  is
fitted by a tight binding model with two Fe ions per unit cell and
two degenerate orbitals per Fe ion. Based on this, superconductivity
with extended s-wave pairing symmetry of the form $\cos k_x +\cos
k_y$ is examined. The local density of states near an impurity is
also investigated by using T-matrix approach. For the nonmagnetic
scattering potential, we found that there exist two major resonances
inside the gap. The height of the resonance peaks depends on the
strength of the impurity potential. These in-gap resonances are
originated in the Andreev's bound states due to the quasiparticle
scattering between the hole Fermi surfaces around $\Gamma$ point
with positive order parameter and the electron Fermi surfaces around
$M$ point with negative order parameter.

\end{abstract}


\maketitle

The recent discovery of a new family of superconductors, i.e. the
FeAs based superconductors [1-6], has attracted much attention in
the condensed matter community. It has been reported that the
superconducting transition temperature $T_c$ can be obtained as high
as 55K [2]. The undoped iron arsenides have a spin density wave
order below 150K [4]. When holes or electrons are doped, the iron
arsenides become superconducting.

Similar to the cuprate superconductors, the FeAs based
superconductors also have a layer structure. It has been accepted
that superconductivity comes from the Cooper pairs in the Fe-Fe
plane. However, in the FeAs based superconductors, each unit cell
contains two Fe ions and two As ions. The four As ions around each
Fe ion do not locate in the Fe-Fe plane and have a two-fold rotation
symmetry and two reflection symmetries (see Fig. 1). Due to
different arrays of As ions around Fe ions, the Fe-Fe plane can be
divided into two sublattices A and B. We note that the diagonal
directions of the Fe-Fe plane have the translational symmetry with
the period $a$. In this coordinate system, the momentum is a good
quantum number.

Angle resolved photoemission spectroscopy (ARPES) experiments have
probed the electronic properties in the FeAs-based superconductors
[7-15]. It is established that there are two hole Fermi surfaces
around (0,0) and two electron Fermi surfaces around $(\pi,\pi)$.
These Fermi surface characteristics have been obtained by the LDA
calculations [16-19]. Many theoretical models have been presented to
reproduce the hole and electron pockets by employing the Fe d and As
p orbitals and the hybridization among them [20-25]. However, there
is no consensus on the superconducting gaps on the Fermi surfaces.
In a series of ARPES, scanning tunneling microscopy (STM)
experiments, and point-contact Andreev reflection spectroscopy
experiments, the order parameter has been interpreted to be nodeless
[7,9-15,26], nodal [27-29], single gap [10,26,29-32], and multiply
gaps [7,11-15].

\begin{figure}
\rotatebox[origin=c]{0}{\includegraphics[angle=0,
           height=3.0in]{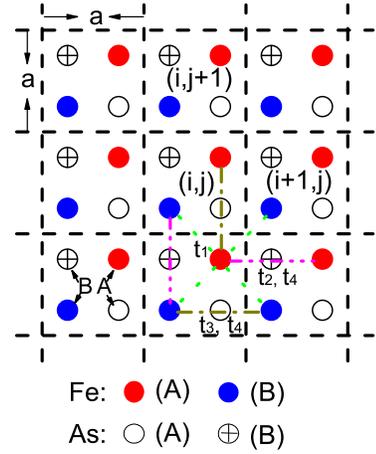}}
\caption{(Color online) Schematic lattice structure of the FeAs
layers with each unit cell (denoted by $i$ and $j$) containing two
Fe (A and B) and two As (A and B) ions. The As ions A and B are
located just above and below the center of each face of the Fe
square lattice, respectively. Here, $t_1$ is the nearest neighboring
hopping between the same orbitals $d_{xz}$ or $d_{yz}$. $t_2$ and
$t_3$ are the next nearest neighboring hoppings between the same
orbitals mediated by the As ions B and A, respectively. $t_4$ is the
next nearest neighboring hopping between the different orbitals.}
\end{figure}

In this work, we start from two Fe ions per unit cell and two
degenerate orbitals $d_{xz}$ and $d_{yz}$ per Fe ion and construct
an effective four-band model, which exhibits the features of the
Fermi surfaces in the FeAs-based superconductors. Based on the mean
field theory for superconductivity, we investigate the differential
conductance and the impurity effect for the extended s-wave pairing
symmetry [12,17,20], so that we can understand the electronic
properties in the FeAs-based superconductors.

We assume that $t_1$ is the hopping between the same orbitals on the
nearest neighboring  Fe sites, $t_2$ and $t_3$ are the next nearest
neighboring hoppings between the same orbitals mediated by As ions B
and A, respectively, and $t_4$ is the hopping between the different
orbitals on the next nearest neighboring Fe sites (see Fig. 1). It
is expected that $t_4$ is small and has the same value in both
translation symmetry directions. Therefore, the model Hamiltonian we
propose can be written as

$$H_0=-\sum_{\alpha ij\sigma }\{\mu({c}_{A\alpha,ij\sigma
}^{+}{c}_{A\alpha,ij\sigma}+{c}_{B\alpha,ij\sigma
}^{+}{c}_{B\alpha,ij\sigma })$$
$$+[t_1{c}_{A\alpha,ij\sigma
}^{+}({c}_{B\alpha,ij\sigma}+
{c}_{B\alpha,i+1j\sigma}+{c}_{B\alpha,ij+1\sigma}+{c}_{B\alpha,i+1j+1\sigma})$$
$$+t_2({c}_{A\alpha,ij\sigma }^{+}{c}_{A\alpha,i+1j\sigma}
+{c}_{B\alpha,ij\sigma }^{+}{c}_{B\alpha,ij+1\sigma })$$
$$+t_3({c}_{A\alpha,ij\sigma}^{+}{c}_{A\alpha,ij+1\sigma}
+{c}_{B\alpha,ij\sigma }^{+}{c}_{B\alpha,i+1j\sigma })$$
$$+t_4({c}_{A\alpha,ij\sigma }^{+}{c}_{A\alpha+1,i+1j\sigma}
+{c}_{A\alpha,ij\sigma}^{+}{c}_{A\alpha+1,ij+1\sigma } $$
$$+{c}_{B\alpha,ij\sigma }^{+}{c}_{B\alpha+1,i+1j\sigma } +
{c}_{B\alpha,ij\sigma }^{+}{c}_{B\alpha+1,ij+1\sigma })+{\rm
h.c.}]\}, \eqno{(1)}$$ where $\sigma$ is the spin index, $i, j$
label the position of unit cell, and $\alpha=0$ and 1 represent the
degenerate orbitals $d_{xz}$ and $d_{yz}$, respectively. Obviously,
$H_0$ possesses the same symmetry with the FeAs-based
superconductors, which is key to understand the electronic
properties of this new family of high temperature superconductors.

\begin{figure}
\rotatebox[origin=c]{0}{\includegraphics[angle=0,
           height=2.4in]{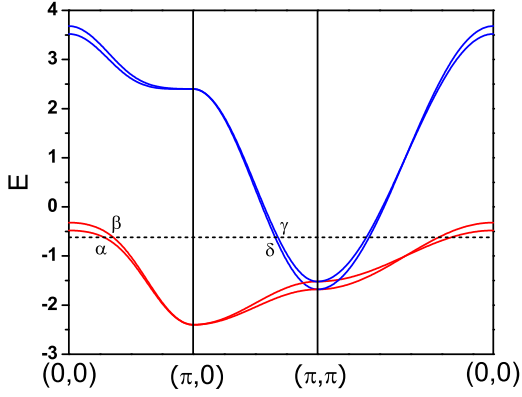}}
\caption {(Color online) The band structure of the four-band model
with $t_1=0.5, t_2=0.2, t_3=-1.0,t_4=0.02$, and $\mu=-0.622$ (eV),
plotted along the path $(0,0)\rightarrow (\pi,0)\rightarrow
(\pi,\pi)\rightarrow (0,0)$.}
\end{figure}

By diagonalizing the above Hamiltonian in the momentum space, one
obtains

$$H_0=\sum_{uv{\bf k}\sigma}(E_{uv,{\bf k}}-\mu)\psi_{uv,{\bf
k}\sigma}^+\psi_{uv,{\bf k}\sigma}, $$
$$E_{uv,{\bf
k}}=\frac{1}{2}(\epsilon_{A,{\bf k}}+\epsilon_{B,{\bf
k}})+(-1)^v\epsilon_{xy,{\bf k}}$$
$$+(-1)^u\sqrt{\frac{1}{4}(\epsilon_{A,{\bf k}}-\epsilon_{B,{\bf
k}})^2+\epsilon_{T,{\bf k}}\epsilon_{T,{\bf k}}^*}, \eqno{(2)}$$
where $u(v)=0, 1$, $\epsilon_{A,{\bf k}}=-2(t_2\cos k_x+t_3\cos
k_y), \epsilon_{B,{\bf k}}=-2(t_2\cos k_y+t_3\cos k_x),
\epsilon_{xy,{\bf k}}=-2t_4(\cos k_x+\cos k_y)$ and
$\epsilon_{T,{\bf k}}=-t_1[1+e^{ik_x}+e^{ik_y}+e^{i(k_x+k_y)}]$.
Here, we have set the lattice constant $a=1$. In deriving Eq. (2),
we have introduced
${c}_{A(B)\alpha,ij\sigma}=\frac{1}{\sqrt{N}}\sum_{\bf
k}{c}_{A(B)\alpha,{\bf k}\sigma}e^{i(k_x x_i+k_y y_j)}$ and have
taken the canonical transformation
$$ \left (\begin{array}{c}c_{A0,{\bf k}\sigma}\\c_{A1,{\bf
k}\sigma}\\c_{B0,{\bf k}\sigma}\\c_{B1,{\bf k}\sigma}
\end{array}\right)=\left (\begin{array}{cccc}\frac{a_{0,{\bf k}}}{\Gamma_{0,{\bf
k}}}&\frac{a_{0,{\bf k}}}{\Gamma_{0,{\bf k}}}&\frac{a_{1,{\bf
k}}}{\Gamma_{1,{\bf k}}}&\frac{a_{1,{\bf k}}}{\Gamma_{1,{\bf k}}}\\
\frac{a_{0,{\bf k}}}{\Gamma_{0,{\bf k}}}&-\frac{a_{0,{\bf
k}}}{\Gamma_{0,{\bf k}}}&\frac{a_{1,{\bf
k}}}{\Gamma_{1,{\bf k}}}&-\frac{a_{1,{\bf k}}}{\Gamma_{1,{\bf k}}}\\
\frac{\epsilon_{T,{\bf k}}^*}{\Gamma_{0,{\bf
k}}}&\frac{\epsilon_{T,{\bf k}}^*}{\Gamma_{0,{\bf
k}}}&\frac{\epsilon_{T,{\bf k}}^*}{\Gamma_{1,{\bf
k}}}&\frac{\epsilon_{T,{\bf
k}}^*}{\Gamma_{1,{\bf k}}}\\
\frac{\epsilon_{T,{\bf k}}^*}{\Gamma_{0,{\bf
k}}}&-\frac{\epsilon_{T,{\bf k}}^*}{\Gamma_{0,{\bf
k}}}&\frac{\epsilon_{T,{\bf k}}^*}{\Gamma_{1,{\bf
k}}}&-\frac{\epsilon_{T,{\bf k}}^*}{\Gamma_{1,{\bf
k}}}\end{array}\right)\left (\begin{array}{c}\psi_{00,{\bf
k}\sigma}\\\psi_{01,{\bf k}\sigma}\\\psi_{10,{\bf
k}\sigma}\\\psi_{11,{\bf k}\sigma}
\end{array}\right),\eqno{(3)}$$
where $\Gamma_{u,{\bf k}}=\sqrt{2(a_{u,{\bf k}}^2+\epsilon_{T,{\bf
k}}\epsilon_{T,{\bf k}}^*)}$ and $a_{u,{\bf
k}}=\frac{1}{2}(\epsilon_{A,{\bf k}}-\epsilon_{B,{\bf
k}})+(-1)^u\sqrt{\frac{1}{4}(\epsilon_{A,{\bf k}}-\epsilon_{B,{\bf
k}})^2+\epsilon_{T,{\bf k}}\epsilon_{T,{\bf k}}^*}$.

Eq. (2) describes analytically  four energy bands with the indexes
$<u,v>$. In Fig. 2, we plot these bands along the path
$(0,0)\rightarrow (\pi,0)\rightarrow (\pi,\pi)\rightarrow (0,0)$. In
our calculations, we have used $t_1=0.5, t_2=0.2,
t_3=-1.0,t_4=0.02$, and $\mu=-0.622$ (half filling) (eV). Obviously,
there exist two hole Fermi surfaces around (0,0), i.e. $\alpha$- and
$\beta$-bands, and two electron Fermi surfaces around $(\pi,\pi)$,
i.e. $\gamma$- and $\delta$-bands. This is consistent with those
observed from ARPES experiments [7-15]. We note that the hole and
electron pockets are associated with $u=1$ and 0 while $v=0$ and 1
represent the inner  and outer Fermi surfaces of the hole and
electron pockets, respectively. The parameters $t_1, t_2$ and $t_3$
determine the sizes of the hole and electron pockets, and $t_4$
controls the intervals between the inner and outer Fermi surfaces.
We note that $\mu<-0.622$ and $\mu>-0.622$ correspond to hole and
electron dopings, respectively. When hole (electron) doping
increases, the hole (electron) Fermi surfaces, i.e. $\alpha$- and
$\beta$-bands ($\gamma$- and $\delta$-bands), become larger while
the electron (hole) Fermi surfaces i.e. $\gamma$- and $\delta$-bands
($\alpha$- and $\beta$- bands) become smaller. The variation of the
Fermi surfaces with hole or electron doping has been observed by
ARPES experiments [7-15]. When $\mu>-0.32$ (i.e. $\sim 26.5\%$
electron doping), $\beta$-band disappears.

In order to investigate superconductivity in iron arsenides, we now
introduce the mean field BCS Hamiltonian
$$H_{SC}=\sum_{uv{\bf k}}(\Delta_{uv,{\bf k}}\psi_{uv,{\bf k}\uparrow}^+\psi_{uv,{\bf
{-k}}\downarrow}^+ +{\rm h.c.}), \eqno{(4)}$$ where $\Delta_{uv,{\bf
k}}$ are the superconducting gaps on the energy bands $<u,v>$,
depending on the momentum of the long-lived quasiparticles
$\psi_{uv,{\bf k}\sigma}$. We can see from Eqs. (3) and (4) that
both inter- and intra-band pairings in the original electron
operators $c_{A\alpha,{\bf k}\sigma}$ and $c_{B\alpha,{\bf
k}\sigma}$ are automatically included. Here, we assume that the
pairing potential between electrons is unique, which can avoid many
superconducting transition temperatures [33]. In other words, the
energy gaps on all the Fermi surfaces can be fitted by a single
function of the momentum, i.e. $\Delta_{10,{\bf k}}=\Delta_{11,{\bf
k}}=\Delta_{00,{\bf k}}=\Delta_{01,{\bf k}}$. In Ref. [12], Nakayama
{\it et al.} measured the energy gaps on different Fermi surfaces in
optimally hole-doped Ba$_{0.6}$K$_{0.4}$Fe$_2$As$_2$ ($T_c \sim
37$K) by employing ARPES experiments. The order parameter can be
fitted as $ \Delta_{uv,{\rm \bf k}}=\frac{1}{2}\Delta_0(\cos
k_x+\cos k_y)$ with $\Delta_0=13.5$ meV or $|\Delta_{uv,{\rm \bf
k}}|$. However, in the STM experiments on optimally electron-doped
BaFe$_{1.8}$Co$_{0.2}$As$_2$ ($T_c\sim 22.5$K) [31,32], only two
coherence peaks were observed at a small gap, i.e. $\sim\pm 5.8$
meV. In the following we shall calculate the differential
conductance for the extended s-wave pairing symmetry in the optimal
electron doping, so that we can compare our theory with the STM
experiments.

\begin{figure}
\rotatebox[origin=c]{0}{\includegraphics[angle=0,
           height=1.5in]{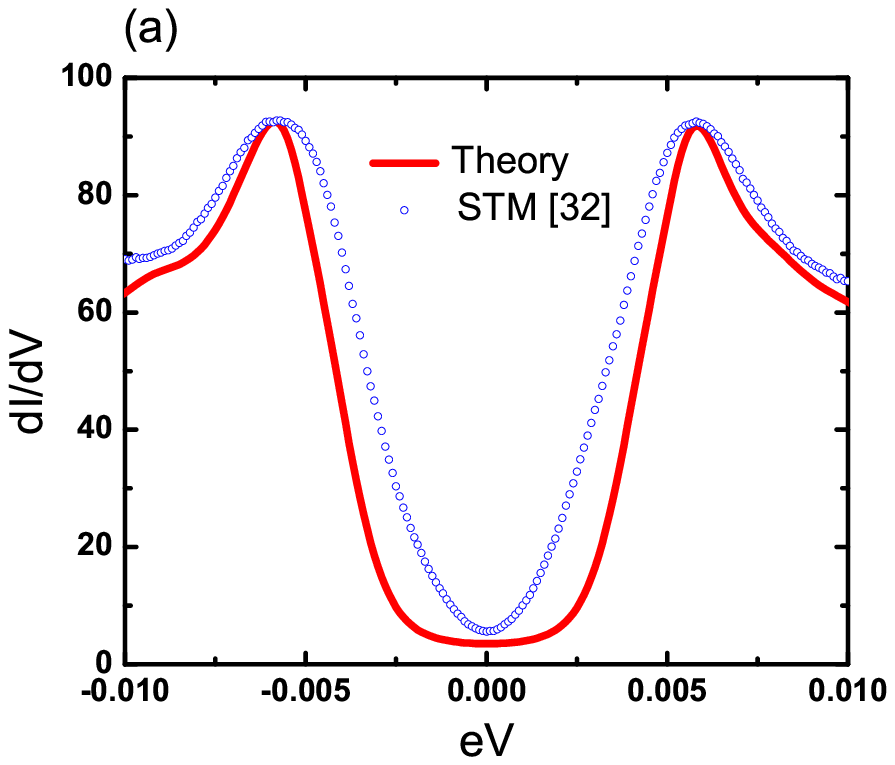}}
\rotatebox[origin=c]{0}{\includegraphics[angle=0,
           height=1.5in]{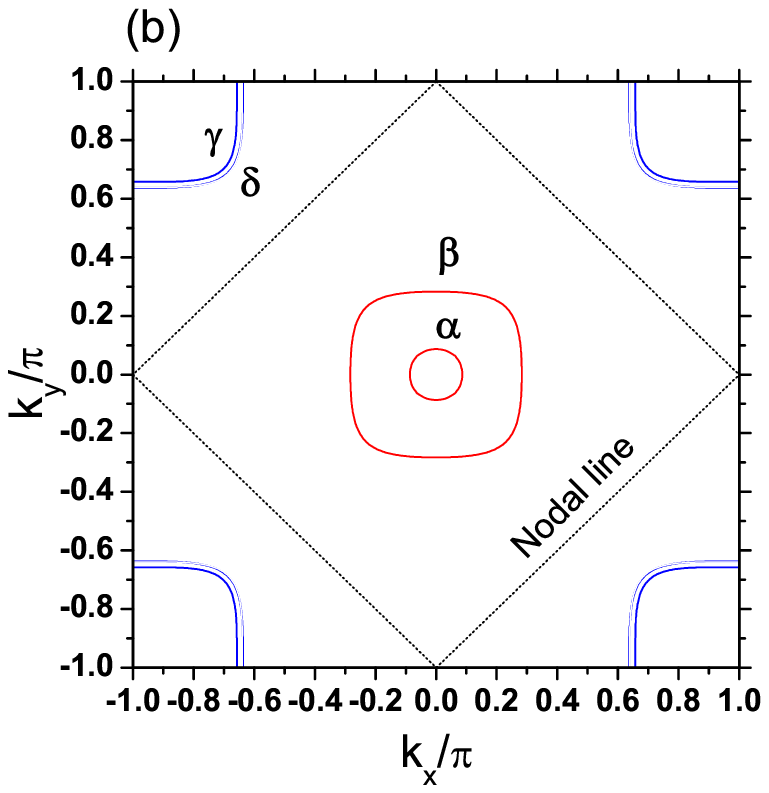}}

\caption {(Color online) (a) The differential conductance  as a
function of the bias voltage eV for the pairing symmetry $
\Delta_{uv,{\rm \bf k}}=\frac{1}{2}\Delta_0(\cos k_x+\cos k_y)$ with
$\Delta_0=5.8$ meV and $|\Delta_{uv,{\rm \bf k}}|$ at optimal
electron doping ($\sim 15\%$) under temperature 4.2K. (b) The
corresponding Fermi surfaces.}
\end{figure}

After diagonalizing the Hamiltonian $H=H_0+H_{SC}$ by the Bogoliubov
transformation, we obtain the local density of states (LDOS) on the
sublattices A or B
$$\rho_{\rm A,B}(\omega)=-\frac{4}{N\pi}\sum_{uv{\bf k}\nu}\frac{{\cal A}^{\rm A,B}_{u,{\bf k}}\xi_{uv,{\bf
k}\nu}^2}{i\omega_n-(-1)^\nu \Omega_{uv,{\bf
k}}}\mid_{i\omega_n\rightarrow \omega+i0^+}, \eqno{(5)} $$ where
$\nu=\pm 1$, ${\cal A}^{\rm A}_{u,{\bf k}}=a_{u,{\bf
k}}^2/\Gamma_{u,{\bf k}}^2$, ${\cal A}^{\rm B}_{u,{\bf
k}}=\epsilon_{T,{\bf k}}\epsilon_{T,{\bf k}}^*/\Gamma_{u,{\bf
k}}^2$, $\Omega_{uv,{\bf k}}= \sqrt{(E_{uv,{\bf k}}-\mu)^2+\Delta
^2_{uv,{\bf k}}}$, and $\xi^2_{uv,{\bf
k}\nu}=\frac{1}{2}[1+(-1)^\nu\frac{E_{uv,{\bf
k}}-\mu}{\Omega_{uv,{\bf k}}}]$. Obviously, the quasiparticles on
the hole and electron pockets have different weights ${\cal A}^{\rm
A,B}_{u,{\bf k}}$ to contribute to the LDOS.

Usually the STM experiments are performed at low temperatures. In
order to compare accurately with the STM experiments, we must take
the effect of temperature into account. The differential conductance
measured by the STM experiments is
$$\frac{dI}{dV}\propto -\int_{-\infty}^\infty f^\prime
(\omega-eV)\rho_{\rm A,B}(\omega)d\omega, \eqno{(6)}$$ where
$f^\prime$ is the derivative of the Fermi function and $V$ is the
bias voltage applied between the STM tip and the sample.

According to the formulas (5) and (6), we can calculate the
differential conductance with different pairing symmetries and
dopings at low temperatures. In Fig. 3(a), we present the
differential conductance for the extended s-wave symmetry with
optimal electron doping under temperature 4.2K. We have observed
that the main contribution to $dI/dV$ comes from the hole Fermi
surfaces, i.e. $\alpha$- and $\beta$-bands. Therefore, whether or
not the nodal points on $\gamma$- and $\delta$-bands exist does not
change qualitatively the features of $dI/dV$. The main difference
between the theoretical results and the STM data could be due to the
fact that either $\beta$- or $\delta$- band of the STM sample is
much closer to the nodal line as depicted in Fig. 3(b).

\begin{figure}
\rotatebox[origin=c]{0}{\includegraphics[angle=0,
           height=1.4in]{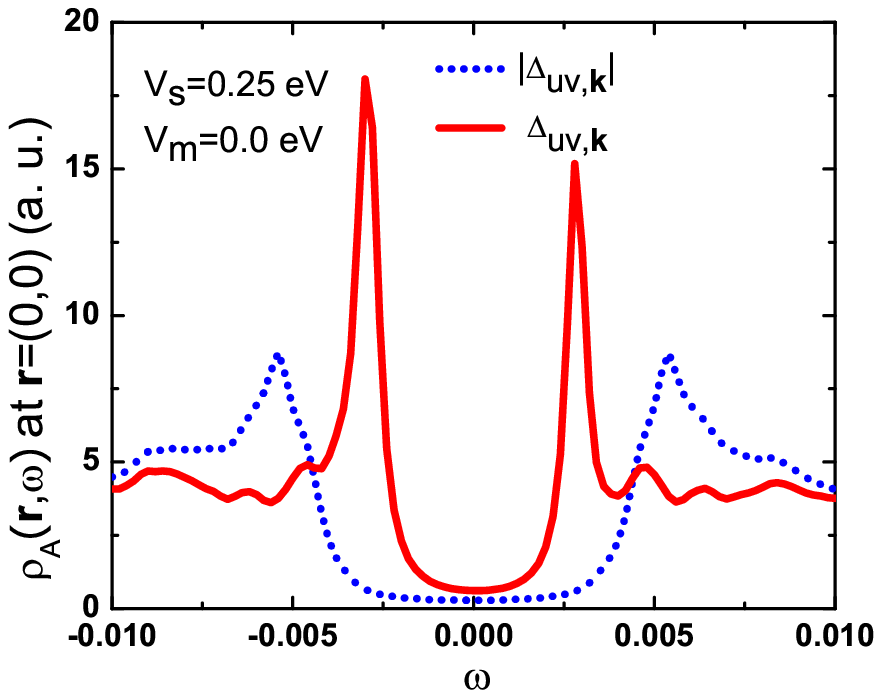}}
\rotatebox[origin=c]{0}{\includegraphics[angle=0,
           height=1.4in]{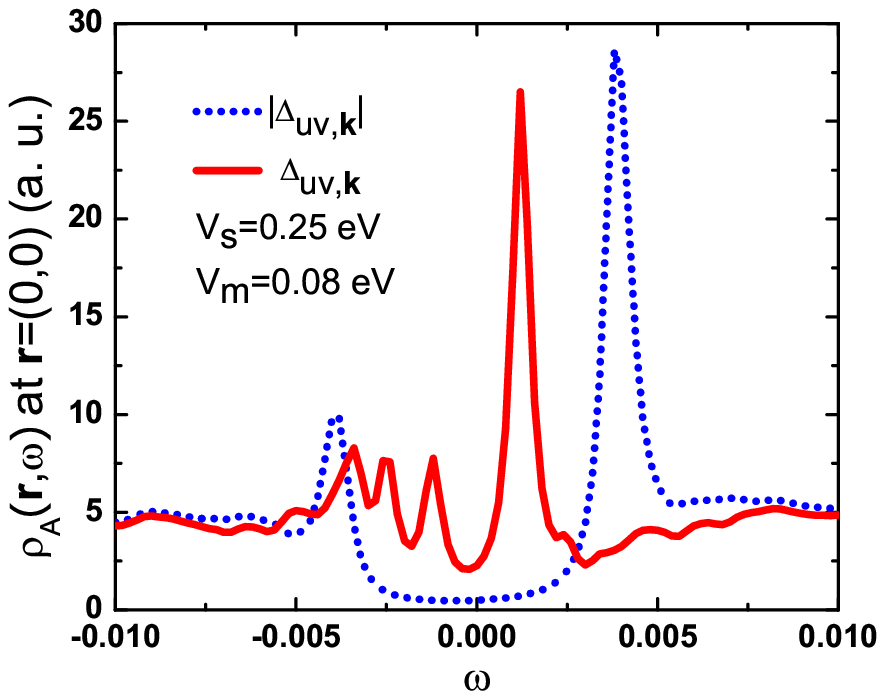}}

\rotatebox[origin=c]{0}{\includegraphics[angle=0,
           height=1.4in]{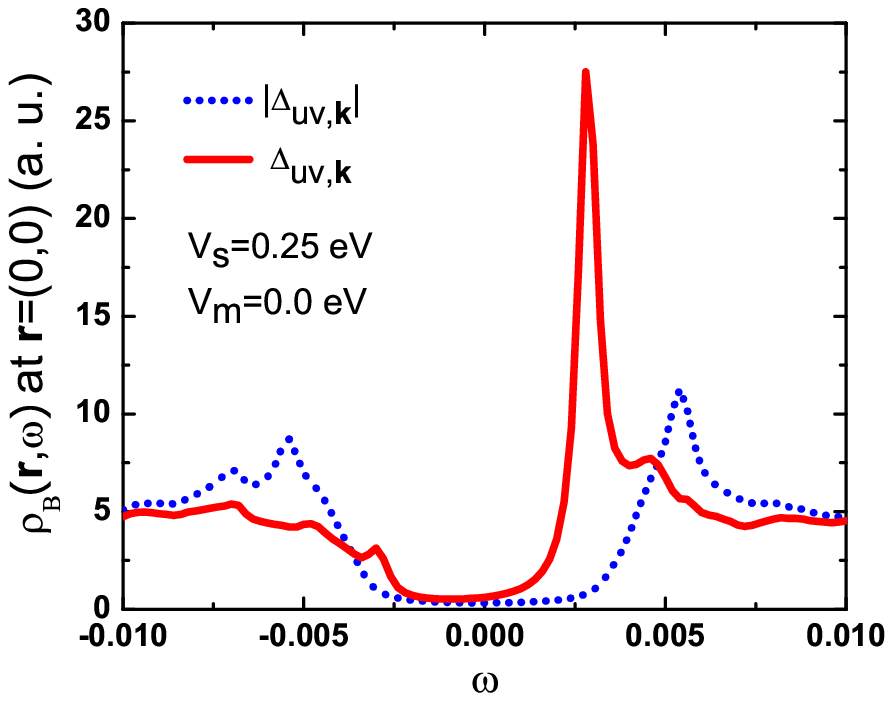}}
\rotatebox[origin=c]{0}{\includegraphics[angle=0,
           height=1.4in]{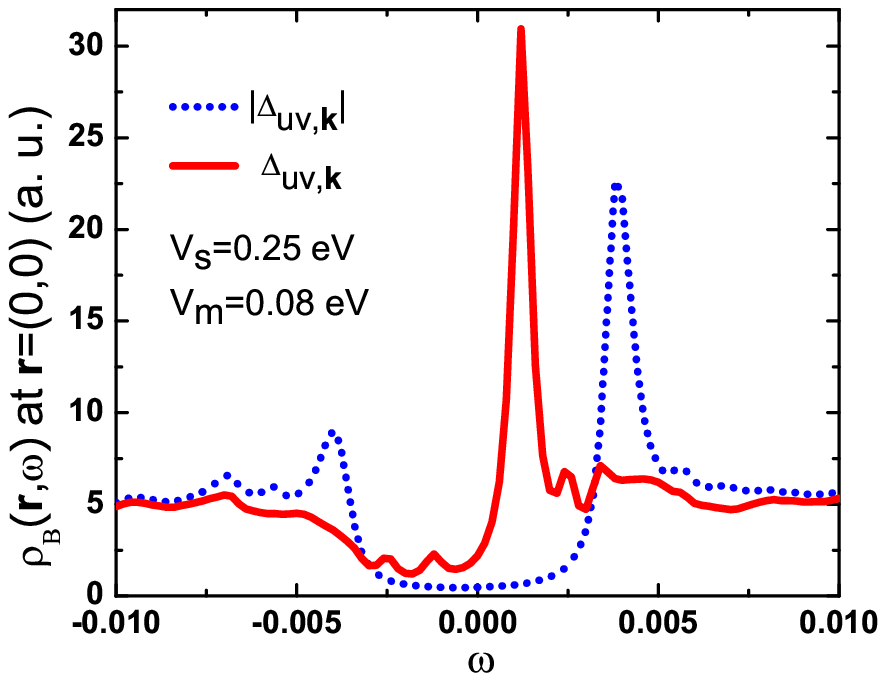}}

\rotatebox[origin=c]{0}{\includegraphics[angle=0,
           height=1.4in]{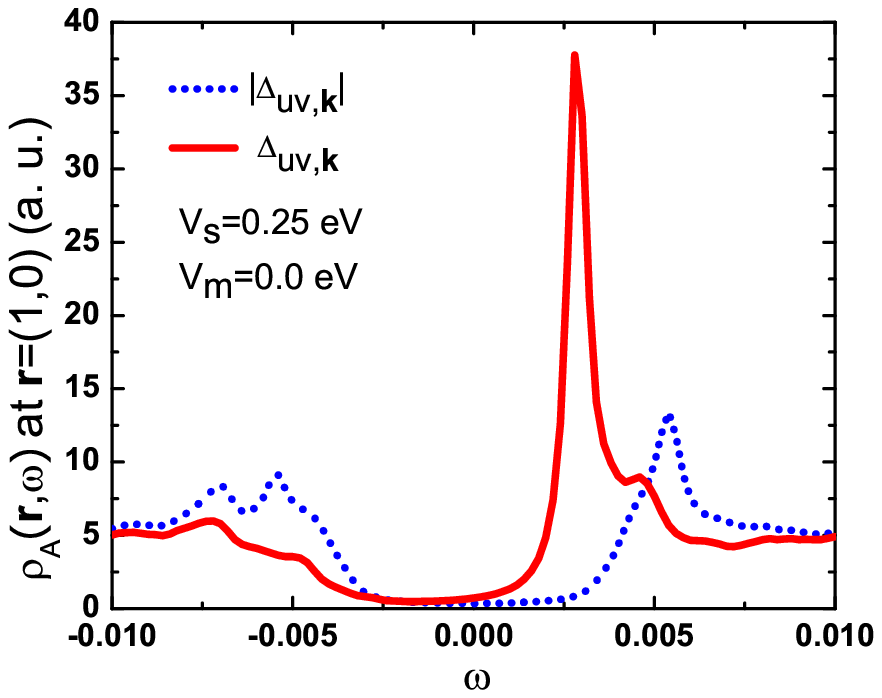}}
\rotatebox[origin=c]{0}{\includegraphics[angle=0,
           height=1.4in]{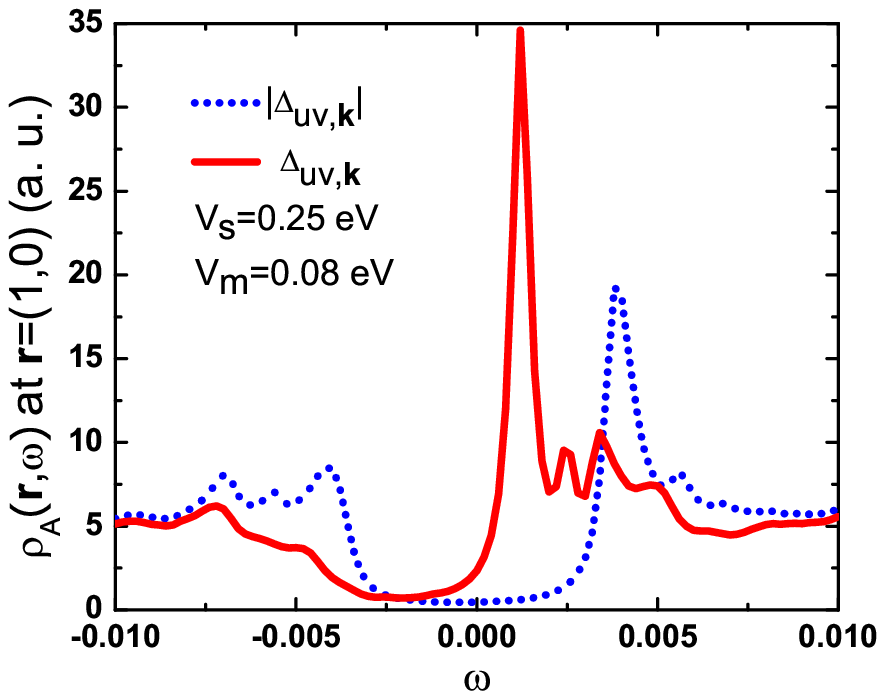}}

\caption {(Color online) The LDOS near an impurity, which includes a
dominant nonmagnetic potential $V_s$ and a small magnetic part
$V_m$, for the pairing symmetry $ \Delta_{uv,{\rm \bf
k}}=\frac{1}{2}\Delta_0(\cos k_x+\cos k_y)$ with $\Delta_0=5.8$ meV
and $|\Delta_{uv,{\rm \bf k}}|$ at optimal electron doping ($\sim
15\%$).}
\end{figure}

In order to detect the sign reversal pairing in the FeAs-based
superconductors, now we calculate the LDOS near an impurity located
at the origin of the sublattice A described by
$H_{{imp}}=V_s\sum_{\alpha \sigma }{c}_{A\alpha,00\sigma
}^{+}{c}_{A\alpha,00\sigma}+V_m\sum_{\alpha}({c}_{A\alpha,00\uparrow
}^{+}{c}_{A\alpha,00\uparrow}-{c}_{A\alpha,00\downarrow
}^{+}{c}_{A\alpha,00\downarrow}).$ Here, $V_s$ and $V_m$ represent
the nonmagnetic part and magnetic part of the impurity potential,
respectively. The total Hamiltonian $H=H_0+H_{SC}+H_{imp}$ can be
solved by T-matrix approach [34]. The analytical expression for the
LDOS on the sublattices A and B near the impurity has been derived
and will be presented elsewhere [35]. We note that the inter-band
scattering is only allowed for those bands with the same index $v$.

In Fig. 4, we plot the LDOS curves for $\Delta_{uv,{\rm \bf k}}$ and
$|\Delta_{uv,{\rm \bf k}}|$ on and near the impurity site with a
moderate strength of nonmagnetic potential, i.e. $V_s=0.25$ eV, plus
a small magnetic potential, i.e. $V_m=0.08$ eV. Obviously, for a
pure scattering potential ($V_m=0$), the LDOS for $\Delta_{uv,{\rm
\bf k}}$ has two impurity resonance peaks at $\pm \omega_0=\pm 2.8$
meV on the impurity site and has a sharp resonance peak at
$\omega_0$ near the impurity site. However, the LDOS for
$|\Delta_{uv,{\rm \bf k}}|$ has no such in-gap impurity states.
Therefore, these impurity resonances can be used to detect the
sign-reversal pairing in the FeAs-based superconductors. The origin
of these impurity resonances comes from the Andreev's bound states
due to the inter-band quasiparticle scattering with the phase
opposite order parameters, similar to that of the zero bias
resonance peak on the Zn impurity in cuprate superconductors. An
additional small magnetic potential can strongly suppress the
impurity peak at $-\omega_0$ and enhance the impurity peak at
$\omega_0$ on the impurity site. Meanwhile, all the resonance peaks
on different sites slowly move forward to zero energy. We note that
for the mixing potential, the LDOS for $|\Delta_{uv,{\rm \bf k}}|$
is similar to that induced by a magnetic impurity in s-wave
superconductors.

\begin{figure}
\rotatebox[origin=c]{0}{\includegraphics[angle=0,
           height=1.4in]{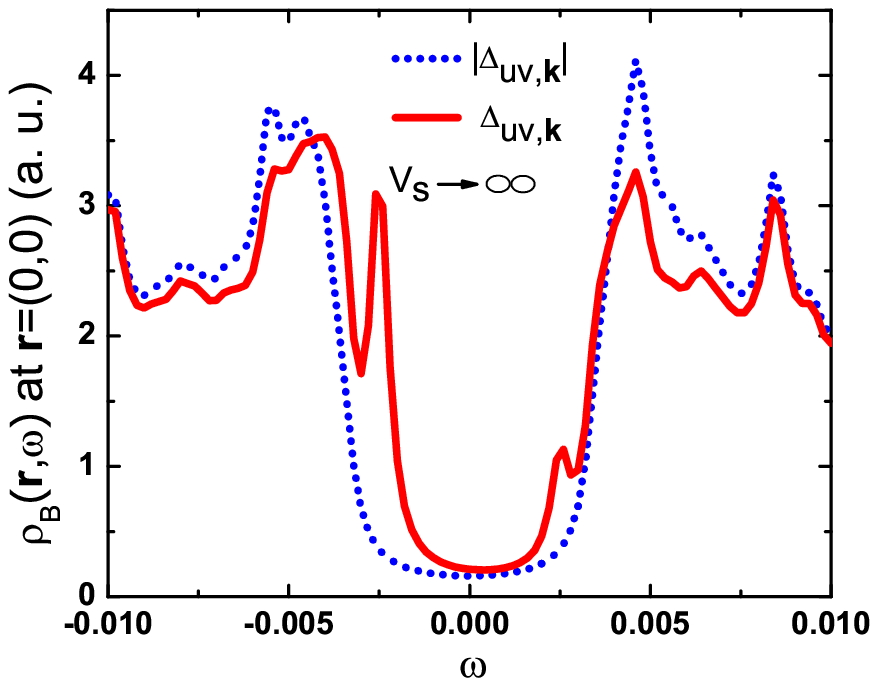}}
\rotatebox[origin=c]{0}{\includegraphics[angle=0,
           height=1.4in]{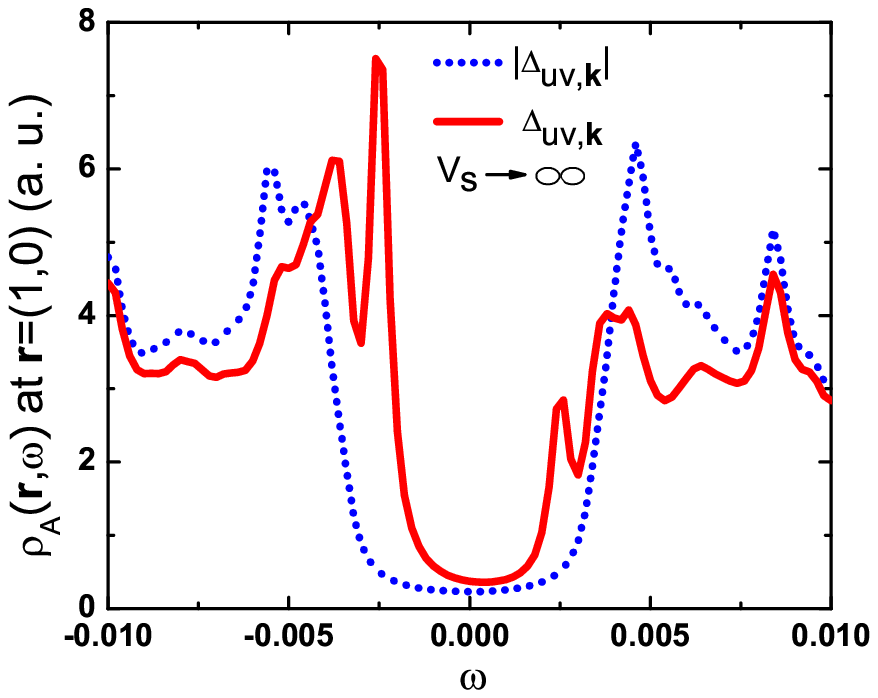}}

\caption {(Color online) The LDOS near a unitary impurity for the
pairing symmetry $ \Delta_{uv,{\rm \bf k}}=\frac{1}{2}\Delta_0(\cos
k_x+\cos k_y)$ with $\Delta_0=5.8$ meV  and $|\Delta_{uv,{\rm \bf
k}}|$ at optimal electron doping ($\sim 15\%$). }
\end{figure}

Fig. 5 shows the LDOS  for $\Delta_{uv,{\rm \bf k}}$ and
$|\Delta_{uv,{\rm \bf k}}|$  near the impurity site with a unitary
potential. The LDOS for $\Delta_{uv,{\rm \bf k}}$ also has two
impurity resonance peaks at $\pm \omega_0$. However, the resonance
peak at $- \omega_0$ is much stronger than that at $\omega_0$. For
$|\Delta_{uv,{\rm \bf k}}|$, the LDOS also has no in-gap impurity
resonance peaks.

We also investigate the other cases of the impurity potential. For
the attractive scattering potential, i.e. $V_s<0$, the stronger
resonance peak inside gap in the LDOS near the impurity site always
appears at negative energy. With increasing $|V_s|$, the resonance
peaks become higher. When $V_s\rightarrow -\infty$, the LDOS is
identical with that for $V_s\rightarrow +\infty$, shown in Fig. 5.
We note that an extra small magnetic potential does not change the
features of the LDOS. For a dominantly magnetic potential, the LDOS
near the impurity site for $ \Delta_{uv,{\rm \bf k}}$ has similar
structures with that for $ |\Delta_{uv,{\rm \bf k}}|$, although the
values of $V_m$ or the locations of in-gap resonance peaks are
different. Therefore, the magnetic impurity seems not to be good
tool to detect the sign reversal pairing in the FeAs-based
superconductors.

In summary, we have built a two-orbital four-band tight-binding
model describing correctly the characteristics of the Fermi surfaces
in the FeAs-based superconductors. In the framework of mean field
theory, we have studied the differential conductance and the
impurity effect for extended s-wave pairing symmetry. It is shown
that the in-gap impurity resonances induced by nonmagnetic
scattering potential can be regarded as a signature of sign-reversal
pairing symmetry in the FeAs-based superconductors, which could be
detected by STM experiments. These resonance peaks also exhibit in
the overdoped and underdoped FeAs-based superconductors.

The author would like to thank C. S. Ting, S. H. Pan, Ang Li, and
Tao Zhou for useful discussions, and especially S. H. Pan and Ang Li
for providing me their STM data. This work was supported by the
Texas Center for Superconductivity at the University of Houston and
by the Robert A. Welch Foundation under the Grant no. E-1411.


\end{document}